\documentclass[usegraphicx]{mn2e}

\usepackage{times, amssymb}

\title[Current membership of NGC~2438 in M46 ruled out]{AAOmega radial velocities 
rule out current membership of the planetary nebula NGC~2438 in the open cluster M46}
\author[Kiss et al.]{L.L. Kiss$^1$\thanks{E-mail: laszlo@physics.usyd.edu.au}, Gy.M.
Szab\'o$^{2,3}$, Z. Balog$^4$, Q.A. Parker$^{5,6}$, and D.J. Frew$^{5,7}$\\
\\
$^1$ Institute of Astronomy, School of Physics A28, University of Sydney, NSW 2006, Australia\\
$^2$ Department of Experimental Physics and Astronomical Observatory, University of Szeged,
Hungary\\
$^3$ Group of Eight European Fellow, University of Sydney, NSW 2006, Australia\\
$^4$ Steward Observatory, University of Arizona, Tucson, AZ85721, USA\\
$^5$ Department of Physics, Macquarie University, Sydney, NSW 2109, Australia\\
$^6$ Anglo-Australian Observatory, Epping, NSW 1710, Australia\\
$^7$ Perth Observatory, Bickley, WA 6076, Australia\\
}

\begin{document}

\date{Accepted ... Received ..; in original form ..}


\maketitle

\begin{abstract}

We present new radial velocity measurements of 586 stars in a one-degree field centered on
the open cluster  M46, and the planetary nebula NGC 2438 located within a nuclear radius of the
cluster. The data are based on medium-resolution optical and near-infrared spectra taken with the
AAOmega spectrograph on the Anglo-Australian Telescope. We find a velocity difference of about 30
km~s$^{-1}$ between the cluster and the nebula, thus removing all ambiguities about the
cluster membership of the planetary nebula caused by contradicting results in the literature.
The line-of-sight velocity dispersion of the cluster is 3.9$\pm$0.3 km~s$^{-1}$, likely to be
affected by a significant population of binary stars.

\end{abstract}


\section{Introduction}

Any physical associations discovered between planetary nebulae (PNe),  the short-lived but
spectacular late evolutionary stage of small and intermediate mass stars (between 1--8
M$_\odot$), and star clusters would be a valuable discovery that provides a means of 
establishing  accurate astrophysical parameters for the nebulae through fixing distances and
progenitor ages from cluster isochrones. Accurate distances are particularly useful, from which
one can infer PNe physical properties such as the absolute magnitude of the central stars,
accurate physical dimensions and fluxes. Also, they would provide excellent calibrators for the
surface brightness--radius relation (Frew \& Parker 2006, Frew 2008). Whereas PNe have been
found in 4 globular clusters of the Milky Way (M15, M22, Pal 6 and NGC~6441; Jacoby et al.
1997), none has been reported in the literature as an unambiguous member of a much younger open
cluster (OC). The interest in the latter case is not only due to being able to determine
independent distances to individual nebulae, but also because in a young open cluster the
progenitor of a now-visible PN will be a reasonably constrained higher mass star than those in
globular clusters. This fact offers the opportunity to calibrate the initial-to-final mass
relation of stars on a broad range of masses, usually done by modelling white dwarf populations
in  open clusters (e.g. Weidemann 2000, Dobbie et al. 2006).

Recently, Majaess, Turner \& Lane (2007; MTL07) and Bonatto, Bica \& Santos Jr (2008; BBS08)
have performed detailed investigations of possible physical associations between PNe and OCs. MTL07 considered the cluster membership for 13 PNe that are located in close
proximity to open clusters lying in their lines of sight and listed another 16 PNe/open cluster
coincidences, which might contain physically associated pairs. However, they noted that we have
yet to establish a single association between a PN and an open cluster based on a correlation
between their full set of physical parameters, including the three key parameters of radial 
velocity, reddening, and distance that need to be in good agreement if an association is to be
viable. BBS08 used near-infrared colour-magnitude diagrams and stellar radial density
profiles to study PN/open cluster association for four pairs. They concluded that the best, but
still only probable, cases are those of NGC~2438/M46 and PK~167$-$01/New Cluster 1.
However, Parker et al. (in prep.) have uncovered another compelling case in an old open cluster
that may prove the best candidate yet for a true OC-PN association.

NGC~2438 is a well-known annular PN located about 8 arcminutes from the core of the bright open
cluster M46 (=NGC~2437; BBS08). Despite its brightness, the cluster was relatively unstudied
until recently (e.g. Cuffey 1941; Stetson 1981). Recently published cluster parameters are
relatively well-determined, e.g. $E(B-V)$=0.10-0.15, D =1.5-1.7 kpc and an age of 220-250 Myr
(Sharma et al. 2006, MTL07, BBS08). The estimated turnoff mass is about 3.5 M$_\odot$ (BBS08).
In addition to the possible association with NGC~2438, M46 is also thought to host the
well-studied post-AGB candidate OH~231.8+4.2 (Jura \& Morris 1985). 

Early studies of the radial velocity of NGC~2438 and M46 (Cuffey 1941; O'Dell 1963) indicated a
difference of $\Delta v_{\rm r}\approx$30 km~s$^{-1}$ between the PN and cluster stars, which
suggested that the pair constitues a spatial coincidence only. Three red giants in the cluster
have systemic velocities (Mermilliod et al. 1989, 2007) identical to that of cluster dwarf
members obtained by Cuffey (1941). However, Pauls \& Kohoutek (1996) rekindled interest in the
possibility of the PN/open cluster association when they found similar velocities for both,
although based on a small number of stars. Both MTL07 and BBS08 pointed out the importance of
measuring sufficient stellar radial velocities for the cluster and the PN to establish if the
proximity is real or only chance superposition. 

Previous distance estimates to the cluster and PN  gave similar results though the distance estimates to
the OC are far more accurate than the crude statistical estimates for the PN distance provided by Zhang
(1995). A distance can be estimated based on the H$\alpha$ surface brightness -- radius (SB-$r$) relation
observed for PNe (e.g. Frew \& Parker 2006; Frew 2008).  Using an updated version of the relation first
presented in Pierce et al. (2004), a distance of 1.4$\pm$0.4 kpc is estimated.  Using instead the relation
applicable to bipolar and bipolar core PNe (Frew, Parker \& Russeil 2006), D = 1.9 $\pm$ 0.5 kpc.  These
values are in broad agreement with other recent SB-r determinations in the radio domain, e.g. Van de Steene
\& Zijlstra (1995), D = 1.7 kpc, Zhang (1995), D = 2.1 kpc, Phillips (2004), D = 1.2 kpc, and
Stanghellini, Shaw \& Villaver (2008), D=1.2 kpc, and are all compatible with the cluster distances
(MTL07, BBS08).

Because of the recently revived interest in possible OC-PN associations, we obtained new
multi-object spectroscopic observations of M46 and NGC~2438, where a putative association has
still not been satisfactorily resolved due to ambiguities and confusion in the literature. Here
we present a statistically significant radial velocity sample of 586 stars within 0.5 degrees of
the cluster centre. The data leave no doubt that the PN is not a member of the cluster.

\section{Observations and data reduction}

\begin{figure}
\begin{center}
\leavevmode
\includegraphics[width=8cm]{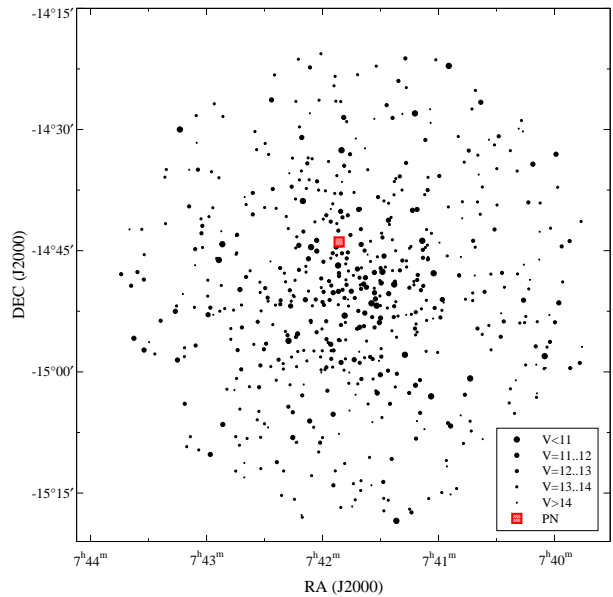}
\end{center}
\caption{Sky positions of the stars observed with AAOmega.}
\label{m46map}
\end{figure}

We used the AAOmega double-beam spectrograph on the Anglo-Australian Telescope in Siding Spring,
Australia on February 17, 2008. The seeing on that night was about 1.5-2$^{\prime\prime}$.  In
the blue arm we used the 2500V grating, providing $\lambda/\Delta \lambda=8000$ spectra between
4800 \AA\ and 5150 \AA. These were important for measuring the mean and expansion velocities of
the PN from the H$\beta$ and [OIII] lines. In the red arm we used the 1700D grating that has
been optimized for recording the Ca~II infrared triplet region. These spectra range from 8350
\AA\ to 8790 \AA, with $\lambda/\Delta \lambda=10000$. This setup has the
highest spectral resolution available with AAOmega and hence the red spectra were used to
measure stellar radial velocities.

\begin{table}
\begin{center}
\caption{\label{obslog} Log of observations.}
\begin{tabular}{lccll}
\hline
Field  & Exp. time & mid-point& Cluster & PN\\
       & (min)           & HJD           &  No. star  & \\
\hline
1      & 40 & 2454513.918   &  282       & N/S rim\\
2      & 65 & 2454513.957   &  304       & central star\\
\hline
\end{tabular}
\end{center}
\end{table}

\begin{figure}
\begin{center}
\leavevmode
\includegraphics[width=7cm]{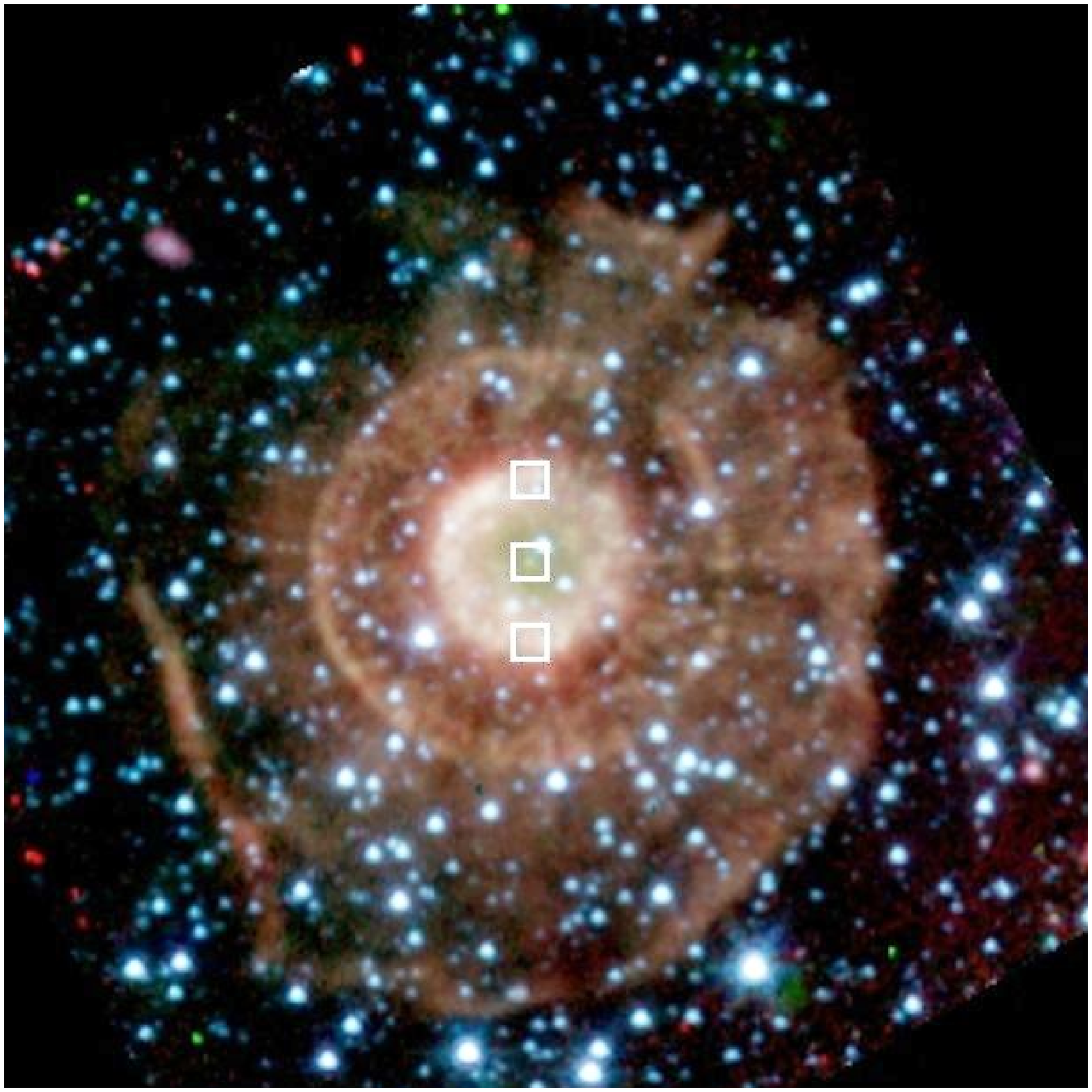}
\vskip1mm
\includegraphics[width=7cm]{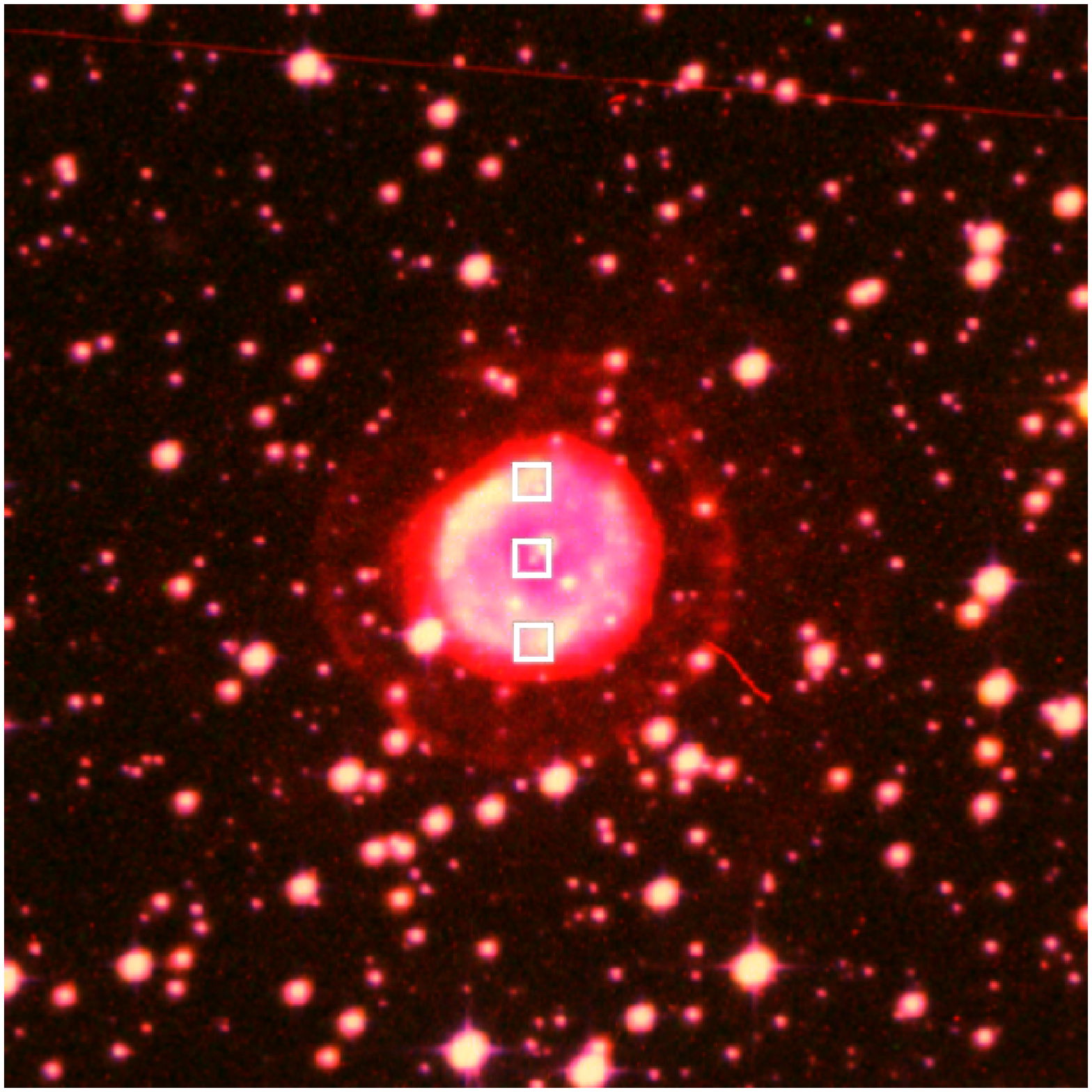}
\end{center}
\caption{Spitzer/IRAC {\it(upper panel)} and SuperCOSMOS H$\alpha$ {\it (lower panel)} images of
NGC~2438. The white squares show the three fiber positions  we used to take spectra of the
planetary nebula. North is up, east is to the left, the field of view is 6 arc minutes.}
\label{pnobs}
\end{figure}

In total, we acquired two field configurations centered on the open cluster. The
target stars were selected from the 2MASS point source catalogue (Skrutskie et al. 2006) by
matching the main features in the colour-magnitude diagram of stars within the central 5 arc
minutes. The total field of view was 1 degree across. We estimated $V$ and $I$ band magnitudes
from the 2MASS $JHK$ magnitudes using the same set of linear transformations as in Kiss et al.
(2007). The full magnitude range of the target stars in $V$ was from 10 mag to 15 mag, but for a
single configuration we limited the brightness range to 3 mag in order to avoid cross-talk
between the fibers due to scattered light. The log of observations is presented in Table\
\ref{obslog}. Fig.\ \ref{m46map} shows the sky positions of the observed stars, while Fig.\
\ref{pnobs} depicts the three fiber positions across the face of the PN (because of the 
limitations of fibre to fibre proximity, the central star was in the first configuration, the
northern and southern rim positions were  in the second configuration).  The upper panel of this
figure is based on Spitzer/IRAC observations of NGC~2438 obtained from the Spitzer archive
(Prog. ID: 68, ``Studying Stellar Ejecta on the Large Scale using SIRTF-IRAC'', PI: G.
Fazio). We downloaded the frames processed with the SSC IRAC Pipeline v14.0, and mosaics were
created from the basic calibrated data (BCD) frames using a custom IDL program. For details see
Gutermuth et al. (2008). As a comparison, the lower panel in Fig.\ \ref{pnobs} shows a deep
H$\alpha$ image (Parker et al. 2005), indicating that the Spitzer-based positions did indeed
coincide with strong optical emission.

The spectra were reduced using the standard 2dF data reduction pipeline. 
We performed continuum normalization separately
for the stellar spectra using the IRAF
task {\it
onedspec.continuum} and then cleaned the strongest skylines that had residuals left using
linear interpolation of the surrounding continuum. The nebular spectra were extracted in
instrumental fluxes only. 

\section{Analysis}

\subsection{Stellar spectra and velocities}

Atmospheric parameters and radial velocity were  determined for each star with an iterative
process, which combined finding  best-fit synthetic spectrum from the Munari et al. (2005)
spectrum library, with $\chi^2$ fitting, and cross-correlating the best-fit model with the
observed spectrum to calculate the radial  velocity. This approach is very similar to that
adopted by the Radial Velocity Experiment (RAVE) project (Steinmetz et al. 2006; Zwitter et al.
2008), and this analysis is based on the very
same synthetic library as RAVE. Our experiences have shown that because of the wide range of
temperatures (and hence spectral features), we needed three subsequent iterations to converge to
a stable set of temperatures, surface gravities, metallicities and radial velocities. The latter
are believed to be accurate within $\pm$1-2 km~s$^{-1}$ for the cooler stars and $\pm$5 
km~s$^{-1}$ for the hotter stars in the sample (the boundary is roughly at 8000-9000 K). These
values have been estimated from Gaussian fits of the cross-correlation profile using the IRAF
task {\it rv.fxcor} and should only be considered as representative numbers. 

\begin{figure}
\begin{center}
\leavevmode
\includegraphics[width=8.5cm]{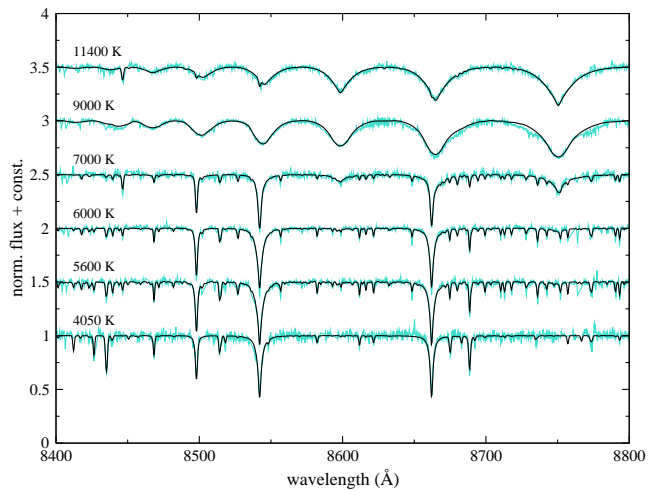}
\end{center}
\caption{Observed stellar spectra (light blue/gray lines) and the best-fit synthetic data from the 
Munari et al. (2005) spectrum library (black lines).}
\label{specfit}
\end{figure}

Although a full discussion of the derived stellar parameters will be presented elsewhere (Kiss
et al., in prep.), we show sample spectrum fits in Fig.\ \ref{specfit} to illustrate the
difficulties one faces when analysing cool and hot stars together in the Ca~II triplet region.
Since the Ca~II lines almost exactly coincide with hydrogen lines in the Paschen series, we
found that it was absolutely crucial to have the best-match template for cross-correlation. A
slightest template mismatch can easily lead to radial velocity shifts of several km~s$^{-1}$ at
this intermediate spectral resolution and hence one has to be very careful to optimize template
selection. It is a commonly used practice in the optical range that the same template is used
across a range of spectral subtypes or even types. However, that does not work in the Ca~II IR
triplet region, where a full $\chi^2$ fit of the spectra is essential. It is also inevitable
that as soon as the temperature reaches about 9000 K, the broad spectral features will lead to a
degraded velocity precision simply because of the broadened cross-correlation profile. M46, as
an intermediate-age open cluster, still hosts a large number of hotter main sequence stars and
that implies the possibility of degraded velocity precision for a significant fraction of stars.
But as we show later, we confirm the $\sim30$ km~s$^{-1}$ velocity difference between the
cluster and the PN, so that the temperature dependent velocity uncertainty does not play a role
in relation to their physical association.

We also measured cross-correlation velocities from the lower resolution blue spectra. These data
were less useful because most of the stars have a broad H$\beta$ line and a few
weak features in the recorded wavelength range, leading to a broad cross-correlation
profile and radial velocities accurate to only about $\pm$10 km~s$^{-1}$.

\subsection{PN spectra and velocities}

\begin{figure}
\begin{center}
\leavevmode
\includegraphics[width=8.5cm]{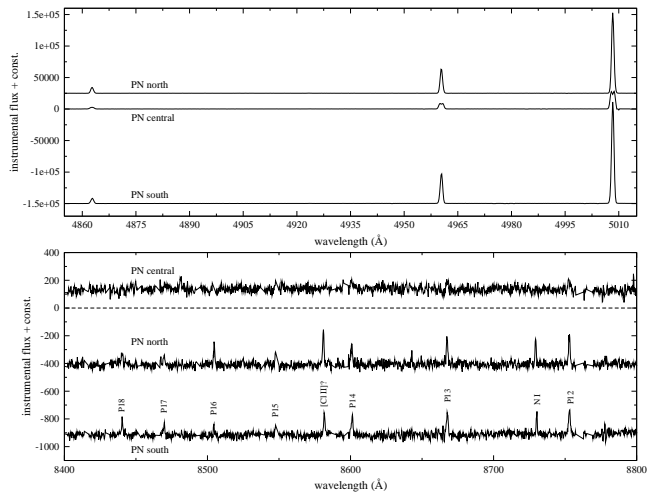}
\end{center}
\caption{{\it Upper panel:} the H$\beta$ and [O III] doublet line profiles in the three observed
positions. The center ones are clearly double-peaked. {\it Lower panel:} the near-infrared
spectra of the same three locations with line identifications. The dashed line shows the zero
flux level for the central star, indicating the detection of a flat continuum. No continuum is
seen in the nebular positions, whose spectra are dominated by the Paschen lines.}
\label{pnspec}
\end{figure}

The spectra taken in the three positions across the PN depicted in Fig.\ \ref{pnobs} are typical
of a planetary nebula. The blue range shows the three characteristic nebular emission lines, 
the H$\beta$ and the [O III] doublet at 4959 \AA\ and 5007 \AA, which are far the strongest
features in the optical spectrum. The central spectrum shows well-defined double-peaked [O III]
line profiles (upper panel of Fig.\ \ref{pnspec}), which we used to determine the expansion
velocity of the nebula, assuming a spherical shell. The two spectra on the edge have
single-peaked emissions, centered exactly halfway between the two peaks of the central spectrum,
supporting that assumption.  

In the near-IR, the central position yielded a featureless flat continuum, while the two
spectra from the shell contain an identical set of narrow emission lines (lower panel in Fig.\
\ref{pnspec}). Using line identifications from the literature (Aller 1977; Rudy et al. 2001) and the NIST atomic database (Ralchenko et al. 2008), we identified the detected lines as
the Paschen series  of hydrogen (from P12 to P18), the [Cl II] nebular line at 8579 \AA, and the
N~I line at 8729 \AA. 

Individual radial velocities have been measured by fitting Gaussian functions to the line
profiles. In case of the double-peaked [O III] doublet, we fitted a sum of two Gaussians. In
each case we repeated the centroid measurement by choosing slightly different fit boundaries to
estimate the uncertainty: the strong emission lines in the blue yielded the same velocities
within 1-2 km~s$^{-1}$ in several repeats, while the weak Paschen lines in the low S/N
near-IR spectra were more sensitive to the actual choice of fitting limits, resulting in
up to 3-5 km~s$^{-1}$ uncertainty per line. 

\section{Discussion}

\subsection{Cluster membership of the planetary nebula}

The Spitzer image of NGC~2438 extracted from the archive is very reminiscent of that of  M57 with
the multiple outer shells of molecular hydrogen seen to extend well ouside the bounds of the well
known optical image. Interestingly here, a deep optical H$\alpha$  image from the SuperCOSMOS
H-alpha Survey (Parker et al. 2005), Fig.\ \ref{pnobs} clearly show faint optical emission too that
matches the inner shell and part of the faint outer shell to  the west seen in the mid-infrared 
(see also fig.\ 7 in Corradi et al. 2003). The Spitzer data also reveal evidence of an interesting
possible interaction of the molecular material with the ISM at the south-eastern edge of the PN,
pointing more or less away from the center of the open cluster.

We present the measured PN emission line velocities in Table\ \ref{pnvr}. The numbers show a very
good agreement between the northern and the southern edges, suggesting a mean velocity of 78$\pm$2
km~s$^{-1}$. The only outlier is the P16~8502 line, however, that feature was significantly
affected by the residual skylines at $\lambda\lambda8504.6-8505.1$ \AA\ and their removal during 
the data reduction. The mean velocity value is also in perfect agreement with the average
mid-point of  the double-peaked [O III] lines in the centre (77.5$\pm$1 km~s$^{-1}$). We therefore
adopt $v_{\rm PN}$=78$\pm$2 km~s$^{-1}$ as the radial velocity of NGC~2438, which is in excellent
agreement with other published velocities in the literature (77 km~s$^{-1}$, Campbell \& Moore
1918;  74$\pm$4 km~s$^{-1}$, Meatheringham, Wood \& Faulkner 1988; 74$\pm$5 km~s$^{-1}$, Corradi et
al. 2000), as well as an unpublished determination of  73$\pm$6 km~s$^{-1}$ from a SAAO 1.9-m
long-slit spectrum (Frew 2008). On the other hand, from the relative velocity difference between
the two peaks in the [O III]  profiles we measured an expansion velocity of 21.0$\pm$0.2
km~s$^{-1}$, virtually identical, for instance, to what Corradi et al. (2000) reported from
high-resolution H$\alpha$ and [O III] spectra (21 km~s$^{-1}$). In summary, our PN measurements
draw a picture that agrees exceptionally well with other results in the literature, confirming the
reliability of the determined velocities and the quoted uncertainties.

\begin{table}
\begin{center}
\caption{\label{pnvr} Measured PN velocities. See the text for a discussion of the 
uncertainties.}
\begin{tabular}{lrrr}
\hline
Line   & PN central & PN north & PN south\\
        & (km~s$^{-1}$) & (km~s$^{-1}$) & (km~s$^{-1}$)\\
\hline
H$\beta$ 4861 &         &   76.1  &  80.1  \\
$[$O III$]$ 4959 & 58.7, 100.2    &  78.6        & 79.8    \\
$[$O III$]$ 5007 &  55.6, 98.2   &  77.6   & 77.6   \\
P16 8502       &       &  68.1        &  58     \\
P15 8545       &       &  75.5      &   76.5      \\
P14 8598       &       &  72.1      &   84.4   \\
P13 8665        &      &  74.3        &  79.5   \\
P12 8750       &       &  73.1      &   76.5   \\
\hline
\end{tabular}
\end{center}
\end{table}

\begin{figure}
\begin{center}
\leavevmode
\includegraphics[width=8cm]{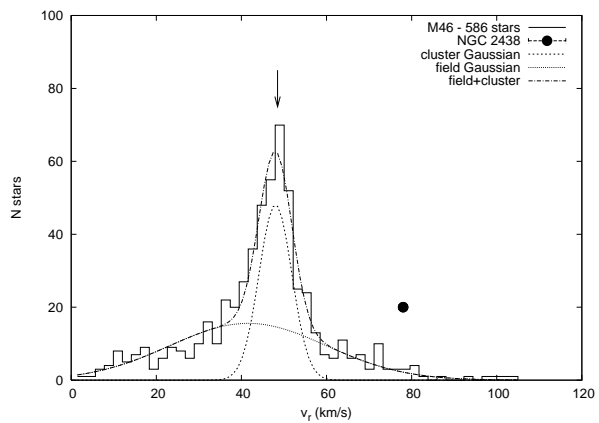}
\end{center}
\caption{The histogram of stellar radial velocities for M46. The arrow shows the mean
center-of-mass velocity of three red giant binaries published by Mermilliod et al. 
(1989, 2007), while the smooth lines represent the Gaussian fits of the field and 
the cluster. About half of
the observed stars belong to the cluster.}
\label{m46hist}
\end{figure}

Fig.\ \ref{m46hist} compares the mean PN velocity to the histogram of the newly derived cluster star
velocities that represent the most extensive and accurate such data for M46 obtained to date. 
We confirm the
early results that NGC~2438 has a relative velocity of about 30 km~s$^{-1}$ with respect to the
cluster (O'Dell 1963), hence the {nebula is not a bound member of the cluster} despite being
located approximately at the same $\sim$1.5-1.7 kpc distance (MTL07, BBS08). In Fig.\ \ref{m46hist}
we also put an arrow at the mean center-of-mass velocity (48.5 km~s$^{-1}$) of three red giant
binaries  measured by Mermilliod et al. (1989, 2007). The excellent agreement between the maximum
of the histogram (49 km~s$^{-1}$ for the highest value, 48 km~s$^{-1}$ for the centroid of the
fitted Gaussian) and the very accurate CORAVEL data confirms both the cluster membership of those
systems and the quoted accuracy of our single-epoch velocity measurements. We also note that while this
paper was under review, a new cluster radial velocity was published by Frinchaboy \& Majewski (2008), who
measured +46.9$\pm$1.0 km~s$^{-1}$ from 19 member stars. The nice agreement gives further support to our
results.

We examined the possibility that the progenitor of NGC~2438 was a runaway star that has been ejected from M46,
since it is the only option for maintaining possible physical association of the cluster and the PN. Runaway
stars escape from star clusters with velocities of typically 30 km~s$^{-1}$ (Blaauw 1961; Hoogerwerf, de
Bruijne \& de Zeeuw 2001) and one cannot ad hoc exclude the possibility that something similar has happened in
M46. However, we feel this to be quite unlikely. If we assume a binary-supernova scenario as a possible
explanation, then the SN must have happened 170-200 Myr ago (i.e. the cluster age minus the evolutionary age
of a progenitor star of 8 M$_\odot$, the lower limit for a core-collapse SN). A star travelling at 30
km~s$^{-1}$ for this time traverses 5-6 kpc in the absence of other forces, so that it should have left the
apparent vicinity of the cluster a long time ago, unless its radial velocity is almost perfectly
aligned along the line-of-sight. The other possibility could be a very recent ejection, most likely within
the last couple of million years, as part of a binary-binary dynamical ejection event. However, the dynamical
ejection scenario (DES) requires a high-density environment in which close encounters of binaries can happen
more frequently: young open clusters or OB  associations forming massive stars (Hoogerwerf, de Bruijne \& de
Zeeuw 2001). M46 is neither young nor has an exceptional high density compared to other intermediate-age open
clusters. Ultimately, future proper motion measurements could be used to distinguish between the option of
chance projection and the option of runaway star. 

In summary, while past membership of the progenitor cannot be completely ruled out, we can safely
conclude against present membership of the PN in the cluster.

\subsection{Evidence of a significant binary population of the cluster}

We also noted the surprisingly large velocity dispersion of the cluster, indicated by the broad
cluster peak of the histogram. While some of it might be explained by the degraded velocimetric
accuracy for the hotter stars (cf. Sect. 3.1), the full range between 35 km~s$^{-1}$ and 60
km~s$^{-1}$ seems to be too large as purely due to measurement errors. We quantified the velocity
dispersion by fitting two Gaussians to the histogram, one broad and one narrow to represent the
galactic background and the cluster members, respectively. The resulting line-of-sight velocity 
dispersion of the cluster is $\sigma_{\rm los}=$3.9$\pm$0.3 km~s$^{-1}$ (removing 78 stars with
$T_{\rm eff}>$10000 K changes the result to $\sigma_{\rm los}=$3.8$\pm$0.2 km~s$^{-1}$). This is a
very high value for a relatively old open cluster and suggests the presence of many binaries.
The velocity dispersion of the broad Gaussian is 18.4 km~s$^{-1}$, which is consistent with field
stars predominantly in the thin disk  (Veltz et al. 2008). 

As a simple exercise, we estimated the dynamical mass of the cluster using the  equation
$M_{\rm dyn}=\eta R_{\rm hl} \sigma_{\rm los}^2 /G,$ where $R_{\rm hl}$ is the half-light
radius, $G$ is the gravitational constant, and $\eta$ is a dimensionless constant (Spitzer
1987). With this we also make the assumption that the cluster is in virial equilibrium, which
is most likely the case, given that clusters older than $\sim$50 Myr are expected to be in virial
equilibrium (e.g. Goodwin \& Bastian 2006).
A rough estimate for the half-light radius of M46 can be taken as the core radius  of
3.3 pc (Sharma et al. 2006). Similarly, about half of the likely members in our spectroscopic
sample lie within 3.4 pc to the cluster centre. Assuming $R_{\rm hl}$=3.3 pc and the canonical
$\eta=9.75$, the result is $M_{\rm dyn}=$1.1$\times$10$^5$ M$_{\odot}$.

On the other hand, adopting $V_{\rm tot}$=6.1 mag from SIMBAD and $V-M_{\rm V}\approx$11.2
mag, the resulting  absolute brightness $M_V=-5.1$ mag corresponds to $L_{\rm
V}/L_\odot\approx$8800, leading to $L_{\rm V}/M_{\rm dyn}\lesssim$0.1, which is way too low for any
stellar cluster (see, e.g., fig. 5 in Goodwin \& Bastian 2006). Therefore, the dynamical mass must
be grossly overestimated, which is a well-known effect when a significant binary population exists
in a cluster (for a recent study see, for instance, Kouwenhoven \& de Grijs 2008). Other evidence
that points in this direction was mentioned by Sharma et al. (2006), who noted that the broad
main-sequence in the colour-magnitude diagram could be due to the presence of binary stars. In that
case orbital motion can introduce extra velocity scatter, which can lead to significantly
overestimated  dynamical masses.

One caveat is the neglected effect of the radial velocity errors on the velocity  dispersion.
If the real errors are much larger than the quoted formal errors from the cross-correlation (cf.
Sect. 3.1), then the velocity dispersion could be dominated by measurement errors, hence getting
only an upper limit to $\sigma_{\rm los}$ and $M_{\rm dyn}$ and a lower limit on $L_{\rm V}/M_{\rm
dyn}$. Our experiences with the same instrument and data on other, mostly globular, star clusters
(Kiss et al. 2007; Sz\'ekely et al. 2007) showed that high S/N AAOmega spectra can reproduce
published radial velocities with the quoted $\pm$1--2 km~s$^{-1}$ accuracy for stars cooler than
$\sim$6000 K. For M46, we have neither repeated observations nor independent measurements with
other instruments to check the precision and the accuracy of the velocities, but with the good S/N
for most of the data, we are confident that the dispersion is not dominated by the measurement
errors.

\section{Summary and future work}

The results of this paper can be summarized as follows:

\begin{enumerate}

\item We have obtained medium-resolution optical and near-infrared spectroscopy for the open cluster M46
and the putative associated planetary nebula NGC~2438. A careful analysis of the data firmly established a
mean radial velocity difference of about 30 km~s$^{-1}$ between the two objects. While the available data
do not rule out past physical association if the star has been ejected, their mutual distance is
currently increasing at least by 30 pc/Myr, hence rejecting present membership of the nebula. 

\item The histogram of the radial velocities has been fitted with a sum of two Gaussians,
representing the smooth galactic field and the cluster. Roughly half of the 586 stars belong to the
cluster, while the rest has a velocity dispersion of about 18.4 km~s$^{-1}$, that is characteristic
for the thin disk. The line-of-sight dispersion of the cluster is 3.9 km~s$^{-1}$, which implies an
unrealistically large dynamical mass. This could be due to the presence of a significant population
of binary stars or larger-than-assumed velocity errors or the combination of both. The large
intrinsic width of the main-sequence in the colour-magnitude diagram supports the binary
explanation. 

\end{enumerate}

Future work should address proper motions of individual stars to investigate if the PN  progenitor
was a runaway star ejected from the cluster. While GAIA will measure very accurate proper motions
from space, there might also be historical photographic plates that are suitable for determining
proper motions. We make all the radial velocity data available through an electronic appendix to
this paper. Combined with multicolour photometric data from the literature (e.g. through the
WEBDA\footnote{\tt http://www.univie.ac.at/webda/} database), these will be used to create an
accurate colour-magnitude diagram of the cluster largely cleaned of the Galactic background and foreground.

\section*{Acknowledgments}

This work is based on observations collected at the Anglo-Australian Observatory, Siding Spring,
Australia. It has been supported by the Australian Research Council, a Group of Eight European
Fellowship and the ``Bolyai J\'anos'' Research Fellowship of the Hungarian Academy of  Sciences
(Gy.M.Sz.), and by NASA through contract 1255094, issued by JPL/Caltech (Z.B.). Z.B. also
received support from Hungarian OTKA Grants TS0498872, T024509, and T049082.

\appendix

\section{The data}

\begin{table}
\begin{center}
\caption{Radial velocities of the observed stars. The full dataset
is available in the online version of the paper.}
\begin{tabular}{ccc}
\hline
RA(J2000)  & DEC(J2000) & v$_{\rm r}$ \\
  (deg)     & (deg)       & (km~s$^{-1}$)\\
\hline
115.63433332 & $-$14.83611944 & 58.4\\
115.57545832 & $-$14.92448333  &50.1\\
115.67374998 &$-$14.90831666  &44.8\\
115.74733332 &$-$14.84368055  &56.1\\
115.88504165 &$-$14.80956111  &39.8\\
115.63974998 &$-$14.89661111  &49.1\\
115.91237498 &$-$14.82234166  &52.8\\
115.84887498 &$-$14.89449166  &45.4\\
115.93391665 &$-$14.79845000  & 38.2\\
115.74662498 &$-$14.88211944 & 40.4\\
115.52187498 &$-$14.86364722 & 20.1\\
115.90645832 &$-$14.93075278 & 47.9\\
115.56870832 &$-$14.86408889 & 54.4\\
\hline
\end{tabular}
\end{center}
\end{table}

\end{document}